\documentclass{article}

\newcommand{\ket}[1]{{|#1\rangle}} \newcommand{\bra}[1]{{\langle#1|}}
\def\braket#1#2{\langle{#1}|{#2}\rangle}

\author{Christof Zalka
\footnote{Department of Combinatorics and Optimization, University of
Waterloo, Waterloo, Ontario, Canada N2L 3G1, e-mail: {\tt
zalka@cacr.math.uwaterloo.ca}}  \and Eleanor Rieffel
\footnote{FX Palo Alto Laboratory, 3400 Hillview Avenue, Palo Alto, CA
94304, USA, e-mail: {\tt rieffel@pal.xerox.com}}}

\title{Quantum operations that cannot be implemented using a small mixed
environment.}

\begin{document}

\maketitle

\begin{abstract}
To implement any quantum operation (a.k.a. ``superoperator'' or ``CP
map'') on a $d$-dimensional quantum system, it is enough to apply a
suitable overall unitary transformation to the system and a
$d^2$-dimensional environment which is initialized in a fixed pure
state. It has been suggested that a $d$-dimensional environment might
be enough if we could initialize the environment in a mixed state of
our choosing. In this note we show with elementary means that certain
explicit quantum operations cannot be realized in this way.  Our
counterexamples map some pure states to pure states, giving strong and
easily manageable conditions on the overall unitary transformation.
Everything works in the more general setting of quantum operations
from $d$-dimensional to $d'$-dimensional spaces, so we place our
counterexamples within this more general framework.
\end{abstract}

\section{Quantum operations}
Quantum operations (see e.g. \cite{schum}) are also known as
``superoperators'', ``superscattering operators'' or ``completely
positive maps'' (``CP maps''). They can be viewed as a generalization
of unitary transformations and are the most general transformations
that can be applied to a quantum system in an unknown (possibly mixed)
state. More precisely, quantum operations are the most general
transformations that can be implemented \underline{deterministically},
thus excluding operations which only succeed with a certain
probability, like those depending on a measurement outcome. Under a
quantum operation pure states are frequently mapped to mixed states.

All quantum operations on a $d$-dimensional system can be implemented 
as the partial trace of a unitary operator acting on the system together
with an auxiliary system (the ``environment''). The question is how small
an environment suffices to implement all possible quantum operations on
a $d$-dimensional system. The answer is easy, and has long been known, for
the case in which the environment is initialized in a pure state. The
answer is more interesting when the environment can be initialized in
a mixed state.  

\section{Initializing the environment in a pure state}
\label{pureEnv}

The environment is initialized in a standard pure state (let's call it
$\ket{0}$). We apply an overall unitary transformation to the system
plus environment (which typically involves interaction between the
two). Finally we are only interested in the (now generally mixed)
state of the system alone, thus we ``trace over'' the
environment. Decoherence is a typical example of such a process. It is
easy to see that quantum operations act linearly on density matrices.

We will think of quantum operations as given in the above way, as
overall unitary transformations on a system together with an
environment. For example, a quantum operation on a single qubit system
can be given by the action of an overall unitary transformation on
basis vectors for the space consisting of the system together with the
environment. As the environment starts out in the state $\ket{0}$, we
need only to give the images of the vectors $\ket{i} \ket{0}$ (where
the first vector corresponds to the system and the second to the
environment):
\begin{eqnarray*}
 \ket{0} \ket{0} \to\ket{0} \ket{\psi_{00}} + \ket{1} \ket{\psi_{01}}
\\ \ket{1} \ket{0} \to \ket{0} \ket{\psi_{10}} + \ket{1}
\ket{\psi_{11}}
\end{eqnarray*}
where the environment vectors $\ket{\psi_{ij}}$ have to be such that
the two right hand sides are orthonormal.


The action of the quantum operation on (density) operators is obtained
by first lifting the above overall unitary transformation to the level
of density operators and then tracing over the environment (using
$tr \ket{\varphi}\bra{\psi}=\braket{\psi}{\varphi}$). Thus
$$ \ket{i}\ket{0}\bra{0}\bra{i'} ~~\stackrel{U}{\to}~~ \sum_{j j'}
\ket{j} \ket{\psi_{ij}} \bra{\psi_{i'j'}} \bra{j'}
~~\stackrel{tr_{env}}{\to}~~ \sum_{j j'} \ket{j} \bra{j'}
\braket{\psi_{i'j'}}{\psi_{ij}}.$$
Since only the inner product of the $\ket{\psi_{ij}}$ matter, we may
assume that $\ket{0} = \ket{\psi_{00}}$. Thus we see that any
quantum operation on a single qubit system can be implemented with
a 4-dimensional environment, as the $\ket{\psi_{ij}}$
span at most a 4-dimensional space. For a $d$-dimensional system we
have $d^2$ environment vectors $\ket{\psi_{ij}}$, so an environment of
that dimension will always do. It is also not hard to show that in
this setting some quantum operations really need an environment of
this size.

\subsection{Initializing the environment in a mixed state}
While it does not matter in which pure state we initialize the
environment, initializing in a mixed state rather than a pure state
does matter.  A mixed state is not unitarily equivalent to a pure
state, rather the unitary equivalence is characterized by the
eigenvalues $p_i$ of the density matrix of the state. The question at
hand is whether we could implement any quantum operation with an
environment smaller than $d^2$ if, for each map we want to implement,
we are allowed to initialize the environment in a mixed state of our
choosing.

Without loss of generality we can assume the initial state of the
environment to be of the form $\rho_{env, initial}=\sum_i p_i \ket{i}
\bra{i}$). Because the environment now is smaller, the overall unitary
transformation has fewer parameters, but more of them matter; it is no
longer sufficient to know only the action on the subspace in which the
environment is in state $\ket{0}$.

A parameter count (including the choice of the $p_i$) suggests that an
environment of the same dimension as the system might just be enough
to implement any quantum operation were we allowed to choose the
initial mixed state of the environment. Thus the conjecture in
\cite{lloyd} (see eq.(2)). We give a more general version of the
parameter count in section \ref{paraCount}.

It has been shown \cite{DiVi} for the qubit case ($d$=2) that this 
conjecture is not true; there are quantum operations which cannot be 
implemented with a mixed single qubit environment. The proof in 
\cite{DiVi} was rather ``brute force'', as it used computer algebra 
to show that a system of equations does not have a solution.
Here we give a simple proof using only elementary means that also 
generalizes to arbitrary dimensions.

\subsection{General framework}
\label{framework}
In general we consider maps between systems of (possibly) different
dimensions $d$ and $d'$. We embed the ``initial'' $d$-dimensional
system in a larger system, apply a unitary operator, and then take the
partial trace over everything except the $d'$-dimensional ``final''
system. Thus physically we start with a system in some (generally
mixed) state and end with a different system in some state. Note that
the dimension can increase or decrease.

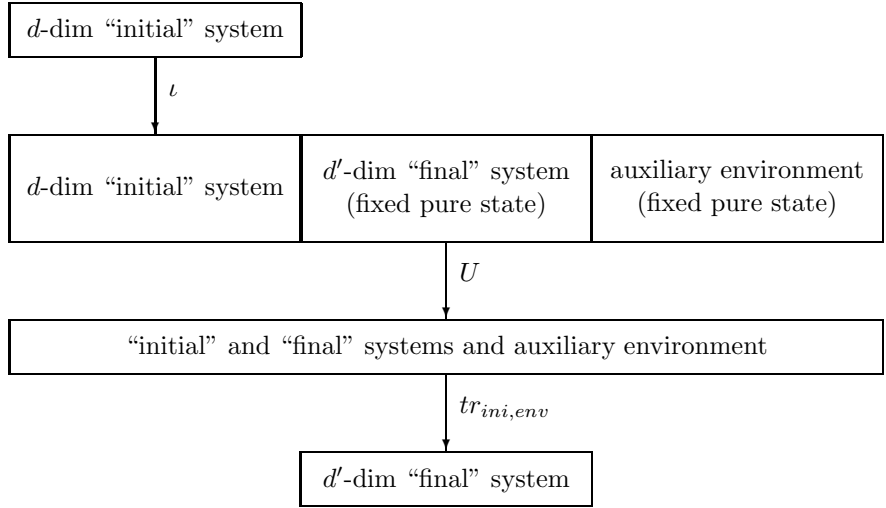
\begin{figure}
\begin{picture}(400, 220)
\put(0,185){\framebox(110, 20){$d$-dim ``initial" system}}
\put(55,185){\vector(0,-1){29}}
\put(60,170){$\iota$}
\put(0,115){\framebox(110, 40){$d$-dim ``initial" system}}
\put(110,115){\framebox(110,40){\shortstack{$d'$-dim ``final" system
 \\(fixed pure state)}}}
\put(220,115){\framebox(110,40){\shortstack{auxiliary environment 
\\(fixed pure state)}}}
\put(165,115){\vector(0,-1){29}}
\put(170,100){$U$}
\put(0,65) {\framebox(330,20){``initial" and ``final" systems and 
auxiliary environment}}
\put(165,65){\vector(0,-1){29}}
\put(170,50){$tr_{ini, env}$}
\put(110,15) {\framebox(110,20){$d'$-dim ``final" system}}
\end{picture}
\caption{Map from $d$-dimensional ``initial" system to $d'$-dimensional
``final" system induced by unitary transformation $U$ acting on the two 
systems and an auxiliary environment. The ``final" system and the
auxiliary environment are  initialized in fixed pure states.}
\label{auxEnv}
\end{figure}

When the larger system contains a $d'$-dimensional system and an
auxiliary environment initialized in pure states (see Figure
\ref{auxEnv}), the same argument as in Section \ref{pureEnv} shows
that besides the $d$-dimensional initial system and the
$d'$-dimensional final system, a $d$-dimensional environment is enough
(and in general necessary) to realize general quantum operations
between the two systems.

\begin{figure}
\begin{picture}(400, 200)
\multiput(0,165)(0,-50) {2}{\framebox(120, 20){$d$-dim ``initial" system}}
\put(60,165){\vector(0,-1){29}}
\put(65,150){$\iota$}
\put(120,115) {\framebox(200,20){$d'$-dim ``final" system (fixed mixed state)}}
\put(160,115){\vector(0,-1){29}}
\put(165,100){$U$}
\put(0,65) {\framebox(320,20){$d +d'$-dim ``initial" and ``final" system}}
\put(220,65){\vector(0,-1){29}}
\put(225,50){$tr_{ini}$}
\put(120,15) {\framebox(200,20){$d'$-dim ``final" system}}
\end{picture}
\caption{General framework: map from $d$-dimensional ``initial" system 
to $d'$-dimensional ``final" system induced by unitary transformation $U$. 
The ``final" system is initialized in a fixed mixed state.}
\label{generalFramework}
\end{figure}
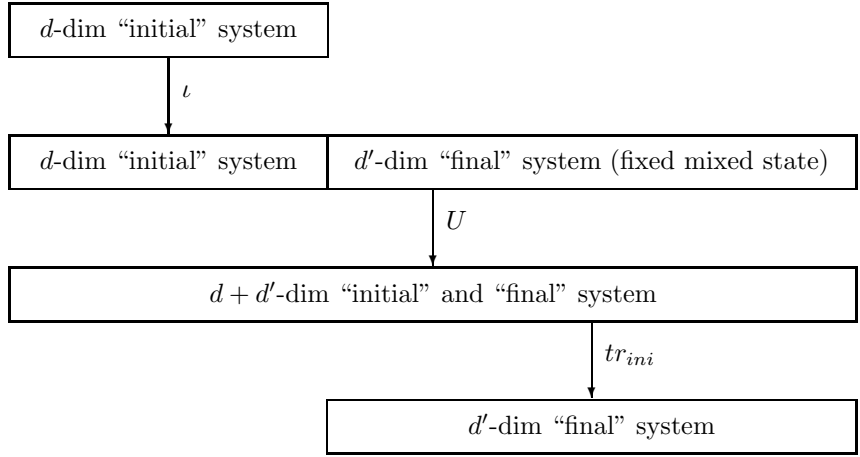

Were we to initialize the ``final'' system in a mixed state, a
parameter count (misleadingly) suggests that the $d$-dimensional
auxiliary environment could simply be left away. We place our
counterexamples in this general framework of maps from a
$d$-dimensional ``initial'' system to a $d'$-dimensional final (or
``target'') system initialized in a mixed state of our choosing. An
overall unitary transformation is applied to the initial plus final
system, and then, as we are only interested in the state of the final
system, we ``trace over'' the initial one. (See Figure
\ref{generalFramework}.)

Note that the question we were initially interested in, of whether
a $d$-dimensional environment that can be initialized in a mixed
state is sufficient to implement any quantum operation on a 
$d$-dimensional system, can be answered in this general framework
by looking at initial and final systems of the same dimension $d$
and then swapping them at the end. 

\subsubsection{A sketch of the parameter count}
\label{paraCount}
Here is a sketch of the parameter count that leads to the wrong
conjecture. The $d'$-dimensional mixed state in the standard form is
given by $d'-1$ (real) parameters. Note that it is invariant under
diagonal (special) unitary transformations (``rephasings''). The
overall (special) unitary transformation has $d^2 d'^2-1$
parameters. Finally we have to take into account that after this
overall transformation, a (special) unitary transformation just on the
initial system alone does not matter. In the end we get
$(d'-1)-(d'-1)+(d^2 d'^2-1)-(d^2-1)=d^2 (d'^2-1)$ real parameters,
which is the correct number for a quantum operation $d \to d'$.

\section{The counterexample}

\subsection{The basic idea}
We explain the basic idea behind the counterexamples in the simple
setting of a map from a $d$-dimensional system to itself induced by a
unitary map on the system together with a $d'$-dimensional
environment.  Imagine we initialize the $d'$-dimensional environment
in a mixed state, say $\rho_{ini} = \rho\ket{0}\bra{0} + (1 -
\rho)\ket{1}\bra{1}$.  The resulting map (quantum operation) will be a
probabilistic mixture (convex linear combination) of the two
``pure-environment" maps we would get by initializing the environment
either in state $\ket{0}$ or $\ket{1}$. A convex linear combination of
quantum operations is defined through the convex linear combination of
the density matrices to which they map. Also note that given one of
these maps, the other has to fulfill certain conditions because they
come from the same overall unitary transformation.

The main idea of our counterexample is that the fact that a pure state 
cannot be written as a non-trivial mixture of two pure states gives us
strong restrictions on maps that map certain pure system states to pure 
states. For example, if a quantum operation maps $\ket{0} \to \ket{0}$,
any ``pure-environment" maps that make up part of a convex linear combination
that gives this map have to map $\ket{0}$ to $\ket{0}$ as well.

\subsection{The counterexample}
As announced, we use the general framework of a quantum operation from
an ``initial'' $d$-dimensional system to a ``final'' $d'$-dimensional
system (Fig. \ref{generalFramework}).  We will mark vectors in the
final system with a prime: $\ket{..}'$.

In our counterexample, we require that all but one of the initial 
system basis states go to one pure state, say $\ket{0}'$, 
$$ \ket{i}\bra{i} \to \ket{0}'\bra{0}' \qquad i=0 \dots d-2. $$
Before further specifying the counterexample, let us look at the
consequences of this condition. If the target system 
has been initialized in a (truly, thus non-pure) mixed state, 
say $\rho'_{ini}=p \ket{0}' \bra{0}'+(1-p) \ket{1}' \bra{1}'$ 
with $0<p<1$, we get for the overall unitary transformation
$$ \ket{i} \ket{0}' \to \ket{\psi_i} \ket{0}' \quad \mbox{and} \quad
\ket{i} \ket{1}' \to \ket{\xi_i} \ket{0}' \qquad i=0 \dots d-2.$$
%
All $2(d-1)$ states on the right hand side have to be orthonormal,
thus the $\ket{\psi_i}$ and $\ket{\xi_i}$ have to be orthonormal. But 
for $d>2$ their number exceeds the dimension of the system. It is clear that
were we to initialize the final system in a mixed state of rank $>2$,
things would only be worse. It follows that the final system
cannot start out in a mixed state, thus we assume it is initialized in
the pure state $\ket{0}'$. (We treat the special case $d=2$ below.)

Without loss of generality we set $\ket{\psi_i}=\ket{i}$. So we have
$\ket{i} \ket{0}' \to \ket{i} \ket{0}'$ for $i=0 \dots d-2$. The image
of the remaining basis state $\ket{d-1}$ can be written
$$\ket{d-1} \ket{0}' \to \sum_{i=0}^{d-1} \ket{i} \ket{\varphi_i}'.$$
Now we require that the image of $\ket{d-1}$ be a truly mixed state
$\rho'_{d-1}$. Furthermore, we require that under the quantum operation, 
$\ket{d-1}$ should ``totally decohere'' from the other basis states 
$\ket{i}$, meaning that a superposition should go to a mixture of the 
individual images of the states, so 
$(\alpha \ket{i}+\beta\ket{d-1})(\bar \alpha \bra{i}+\bar \beta \bra{d-1})
\to |\alpha|^2 \ket{0}' \bra{0}' + |\beta|^2 \rho'_{d-1}
~~(i=0 \dots d-2)$. But the ``totally decohere'' condition means that
$\ket{\varphi_i}'=0$ for all $i \not=d-1$, from which it follows that
the image of $\ket{d-1}$ would have to be pure after all.

\subsection{The special case $d \to d'$ with $d=2$}
For $d=2$ we still have to consider the case in which the final system is
initialized in a mixed state, although from above it is clear that it
would have to be a mixed state of rank 2, thus a mixture of just 2
different pure states. Thus $\rho'_{ini}=p \ket{0}' \bra{0}'+(1-p)
\ket{1}' \bra{1}'$ with $0<p<1$. Then we have for the overall unitary
transformation
$$ \ket{0} \ket{0}' \to \ket{0} \ket{0}' \quad \mbox{and} \quad
\ket{0} \ket{1}' \to \ket{1} \ket{0}'.$$
It follows that $\ket{1} \ket{0}' \to \alpha \ket{0}
\ket{0^\perp}' + \beta \ket{1} \ket{0'^\perp}'$ where both
$\ket{0^\perp}'$ and $\ket{0'^\perp}'$ are vectors orthogonal to
$\ket{0}'$. Similarly for the image of $\ket{1} \ket{1}'$. We can now
make our counterexample above work also for $d=2$ by additionally
requiring that the (truly mixed) image of $\ket{d-1}$ have some overlap with
$\ket{0}'$, thus we require that
$$ \bra{0}' \rho'_{d-1} \ket{0}' \not = 0.$$

\subsection{Summary of the counterexample}
In summary we give the counterexample map by its action on pure
states: 

$$\left(\sum_{i=0}^{d-2} \alpha_i \ket{i} + \beta \ket{d-1}\right) 
\left(\sum_{i=0}^{d-2} \bar \alpha_i \bra{i} + \bar \beta
\bra{d-1}\right) ~~\to~~ 
\left(\sum_{i=0}^{d-2} |\alpha_i|^2 \right)
\ket{0}'\bra{0}'+|\beta|^2 \rho'_{d-1},$$
where $\rho'_{d-1}$ is not a pure state. For $d=2$ we must
additionally require that
$\bra{0}'\rho'_{d-1}\ket{0}'\not=0$. (Actually for $d'=2$ this is
always true.)

\subsection{The counterexample is a quantum operation}
Finally we show that the counterexample is a possible quantum
operation, thus that with a large enough environment it can be
implemented. So besides the initial and final systems we also have an
environment. The map is simple, but for completeness we show it. The
following overall unitary transformation implements the counterexample
map (including the additional property required for the case $d=2$),
with $\alpha,\beta \not=0$:

\begin{eqnarray*}
\ket{i} \ket{0}' \ket{0}_{env} &\to& \ket{i} \ket{0}' \ket{0}_{env}
\qquad i=0 \dots d-2 \\ \ket{d-1} \ket{0}' \ket{0}_{env} &\to& \alpha
\ket{d-1} \ket{0}' \ket{0}_{env} + \beta \ket{d-1} \ket{1}'
\ket{1}_{env}
\end{eqnarray*}

\section{Further remarks}

\subsection{How many counterexamples are there?}
Once a single counterexample has been found, basic topological
properties of the (convex) set of all quantum operations and the (also
convex) subset of those that can be realized with a $d$-dimensional
(possibly) mixed environment imply that the set of counterexamples
contains balls and thus is a set of non-zero measure. 
All that is needed is that the subset of maps that that can be
realized with a $d$-dimensional mixed environment, being convex, is
closed, and that the ``outer'' set is a regular set, thus it is a
closure of its inner points. The latter is simply because the set of
all quantum operations is a ``voluminous'' convex set, thus one which
does not happen to lie in some hyperplane.


Note that our counterexample lies on the boundary of the set of
quantum operations, so we must take the intersection of a ball around
it with the set of quantum operations to find inner points of the set
of counterexamples. Around these points there will then be balls of
counterexamples.

Also note that the counterexample given in \cite{DiVi}, for the single
qubit case ($d=2$), is unital and thus very different from ours.

A complete characterization of exactly which maps can be realized with
a $d$-dimensional mixed environment and which cannot has yet to be found
even in the single qubit case. Ruskai et al. \cite{ruskai} give a 
parameterization of the space of all quantum operations for the single
qubit space that might be useful here. 

%

\subsection{How big an environment is necessary?}
From the counterexamples for $d>2$ we can see that more could be said
about how large a mixed environment has to be, but we haven't
investigated further.

\subsection{Implications for simulating quantum operations (?)}
It looks like our subject is mostly a mathematical challenge. In
\cite{DiVi} it was said that one may want to simulate quantum
operations (e.g. on a quantum computer) with as few resources as
possible (see also \cite{childs}). But Choi \cite{choi} shows that
every extremal map of the convex set of quantum operations can be
realized with a $d$-dimensional environment initialized in a pure
state. To achieve any mixture of such extremal maps (and thus any map)
one can simply carry out each of them with a certain probability,
using a probabilistic protocol. On the other hand certain extremal
maps do need an environment of dimension $d$, thus this probably is a
necessary resource.

\subsection{What about ``Markovian'' quantum operations?}
By ``Markovian'' quantum operations we mean those quantum operations
that can be ``generated'' from ones that are infinitesimally close
to the identity. Of course this notion makes sense only for dimension
preserving quantum operations. It is known that not all quantum
operations can be generated in this way, so we may wonder whether all 
those maps could be implemented with a ``small'' mixed environment.

\paragraph{Acknowledgments}
Part of this work was done during a 3-month stay in the summer of 1999
of Ch.Z. at the FX Palo Alto Laboratory.


\begin{thebibliography}{10}

\bibitem{DiVi}  B.M. Terhal, I.L. Chuang, D.P. DiVincenzo, M. Grassl,
J.A. Smolin \\ {\it Simulating quantum operations with mixed environments}
\\ Phys.Rev. A60 (1999) 881 \qquad (also quant-ph/9806095)

\bibitem{lloyd}S.~Lloyd, {\it Universal Quantum Simulators} \\
Science {\bf 273}, 1073 (1996)

\bibitem{schum} B.~Schumacher, {\it Sending entanglement through noisy
quantum channels} \\ 
Phys.~Rev.~A {\bf 54}, 2614 (1996) \qquad (also quant-ph/9604023)

\bibitem{choi}M.-D.~Choi, {\it Completely positive linear maps on complex
matrices} \\ 
Lin. Algebra and Its Appl. {\bf 10}, 285 (1975).

\bibitem{childs} Dave Bacon, Andrew M. Childs, Isaac L. Chuang, Julia
Kempe, Debbie Leung, Xinlan Zhou, 
{\it Universal simulation of Markovian quantum dynamics} \\ quant-ph/0008070

\bibitem{ruskai} M. B. Ruskai, S. Szarek, E. Werner, {\it An Analysis 
of Completely-Positive Trace-Preserving Maps on 2x2 Matrices}, preprint.
(quant-ph/0101003).

\end{thebibliography}
\end{document}